\documentclass[prd,aps,nofootinbib]{revtex4}
\usepackage[]{epsfig}
\usepackage[centertags]{amsmath}
\usepackage{amsfonts}
\usepackage{amssymb}
\usepackage[english]{babel}

\begin{document}


\title{\bf Spherically symmetric monopoles in noncommutative space}
\author{E. F. Moreno\thanks{Associated with CONICET}}
\affiliation{Departamento de F\'\i sica, Facultad de Ciencias
Exactas, Universidad Nacional de La Plata, C.C. 67, 1900 La Plata,
Argentina\\}
\affiliation{Department of Physics,West Virginia University,
Morgantown, West Virginia 26506-6315, U.S.A.}

\date{\today}

\begin{abstract}
We construct a spherically symmetric noncommutative space in three
dimensions by foliating the space with concentric fuzzy spheres.
We show how to construct a gauge theory in this space and in
particular we derive the noncommutative version of a
Yang-Mills-Higgs theory. We find numerical monopole solutions of
the equations of motion.
\end{abstract}

\maketitle

\section{Introduction}

Field theories in noncommutative space have received renewed
interest since their emergence in certain low-energy limits of
string and M-theory. In particular, non-perturbative soliton
configurations have been the object of numerous investigations in
recent years \cite{NS}-\cite{Correa:2004xm}. Concerning monopoles,
they have been constructed mainly using an extension of Nahm
equation in noncommutative space \cite{Gross1},
\cite{Hamanaka:2001dr}. Other approach exploits the connection
between soliton solutions in four dimensions and monopole
configurations defined in a curved space \cite{Correa:2004xm}.
However, in all of these approaches the obtained configurations
are not the natural extension of the well known 't Hooft-Polyakov
monopole solution. The reason is simple, the 't Hooft ansatz has
explicit spherical symmetry while the standard noncommutative
three dimensional algebra
\begin{equation}
[x_i,x_j]= i \, \theta_{ij} \, , \;\;\;\;\; \theta_{ij}:\,
\text{constant matrix}\label{intro-1}
\end{equation}
is not invariant under rotations, thus breaking explicitly the
rotationally symmetry of any noncommutative field theory (NCFT)
defined on it.

In this article we will construct an explicit rotationally
invariant noncommutative space by deforming adequately the algebra
\eqref{intro-1}. In particular we will show how to construct a
gauge theory in this space by a extending the commutative-space
theory written in terms of explicit rotationally invariant
operators. The evident advantage is that in this formulation the
equations of motion accept a spherically symmetric ansatz,
resemblant to the 't Hooft form. Moreover, we will show that in
the small $\theta$ limit the solutions tends to the well-known
Prasad-Sommerfield solutions.

The article is organized as follows. In section 2 we construct a
rotationally invariant noncommutative space. We find that this
deformation reduces to a foliation of the three-dimensional space
with concentric 2-fuzzy spheres. In section 3 we show how to
construct gauge fields in a manner consistent with the
rotationally symmetry of the space. In section 4 we construct a
Yang-Mills-Higgs theory and derive the equations of motion.
Section 5 is devoted to the solution of the equations of motion.
Finally in section 6 we summarize the paper and present some
discussion.


\section{Rotationally invariant noncommutative space}

One of the main problems in finding noncommutative monopole
solutions is that the simplest ansatz ('t Hooft) has explicit
spherical symmetry whereas the standard noncommutative space in
three dimensions  breaks rotational invariance. Of course,
spherical symmetry is not essential for the construction of
monopole solutions and in fact several non-spherically symmetric
solution has been found explicitly \cite{Gross1},
\cite{Hamanaka:2001dr}, \cite{Correa:2004xm}. However spherical
symmetry greatly simplify the equations by reducing the number of
degrees of freedom. So in order to the take advantage of this
simplification let us modify the noncommutative structure of the
space in order to preserve rotational symmetry.

Consider a three-dimensional noncommutative space with coordinates
satisfying the commutator algebra
\begin{equation}
[x_i, x_j] = i \theta\, \varepsilon_{i j k}\, f(r)\, x_k
\end{equation}
with $f(r)$ a function to be determined and $r^2 = x_i x_i$.

It can be shown that the Jacobi identity imposes the condition
$f(r)\propto r$ ($r^2$ is a Casimir of the algebra), so we have
\begin{equation}
[x_i, x_j] = i \theta\, r\, \varepsilon_{i j k}\, x_k
\label{x-algebra}
\end{equation}
with $\theta$ a dimensionless parameter (here, unlike fuzzy-sphere
coordinates, the coordinates $x_1, x_2,x_3$ are all independent,
there is no constraint between them). Then the operators $x_i/(r
\theta)$ satisfy the $SU(2)$ algebra. Being the algebra
(\ref{x-algebra}) invariant under space rotations, it is natural
to extend it with angular momentum operators $L_i$
\begin{align}
[L_i, L_j] &= i\, \varepsilon_{i j k}\, L_k \nonumber\\
[L_i, x_j] &= i\, \varepsilon_{i j k}\, x_k \label{lx-algebra}
\end{align}
We can find a representation of (\ref{x-algebra}) and
(\ref{lx-algebra}) by identifying the coordinate operators with
$\theta r L_i$:
\begin{equation}
x_i= \theta r L_i
\label{xtol}
\end{equation}
with $L_i$, $SU(2)$ operators. We have that $1/\theta^2 = L_i L_i$
and if restrict ourselves to finite-dimensional representations we
have $1/\theta^2= l(l+1), \; \; l \in \frac{1}{2} \mathbb{N}$ (in
principle we allow spinor representations). In this representation
$r$ is a continuos commutative variable. Notice that the algebra
(\ref{x-algebra})-(\ref{lx-algebra}), for fixed $r$ describes a
fuzzy sphere \cite{Madore:1991rr}-\cite{Abe:2002in}, so
essentially what we are doing is foliating the three-dimensional
noncommutative space with concentric fuzzy spheres.

Since algebra (\ref{x-algebra}) is not invariant under space
translations, is imposible to define momentum operators satisfying
\begin{eqnarray}
%
~[P_i, x_j] &=& -i\, \delta_{ij} \label{px-algebra}
\end{eqnarray}
These relations violate the Jacobi identity for three operators
$\{P_i, x_j, x_k\}$. This is analogous to the fact that for
constant non-commutative space
\begin{equation}
[x_i, x_j] = i \theta_{ij}
\label{standardNC}
\end{equation}
it is not possible to define angular momentum operators satisfying
(\ref{lx-algebra})  since the algebra (\ref{standardNC}) is not
rotationally invariant (the Jacobi identity fails for the triplet
$\{L_i, x_j, x_k\}$).

In order to define a field theory in this noncommutative space we
first define transversal and radial field components and write the
appropriate lagrangian. In commutative space, given a vector field
in Cartesian coordinates $V_i\,,\; i=1,2,3$, we can define
transversal components $V^T_i$ and a radial component $V_r$ as
\begin{align}
V^T_i &= \varepsilon_{ijk}\, x_j\, V_k \nonumber \\
V_r &= -x_i\, V_i \label{transversal-fields}%
\end{align}
The transversal part satisfies the constraint
\begin{equation}
x_i V^T_i = 0
\end{equation}
Cartesian coordinates can be recovered from the transversal and
radial ones through the identity
\begin{equation}
r^2\, V_i = - \varepsilon_{ijk}\, x_j\, V^T_k - x_i\, V_r
\label{cartesian}
\end{equation}
Since we are working in a non-commutative space with explicit
rotational invariance (\ref{x-algebra}) it is natural consider the
transversal and radial fields (\ref{transversal-fields}) as our
primary fields and not the Cartesian components $V_i$. This is
crucial because in noncommutative space there is no mapping as
(\ref{cartesian}) to define Cartesian coordinates.

So, a transversal field in the noncommutative space is a field
that satisfies the constraint $x_i  V^T_i + V^T_i x_i= 0$, or in
virtue of representation (\ref{xtol})
\begin{equation}
L_i  V^T_i + V^T_i L_i= 0 \label{cons0}
\end{equation}
Is straightforward to check that any vector field of the form
\begin{equation}
[L_i, \Phi]
\end{equation}
is transversal. We will see that a slight modification has to be
done in the case of a gauge theory.


\section{Gauge fields}
As we did for arbitrary vector fields we define transversal and
radial gauge fields $A^T_i$ and $A_r$ in analogy with their
commutative counterparts (from now on we will drop the superscript
$T$ in $A^T$). That is $A_i$ and $A_r$ are fields that in
commutative space take the form
\begin{align}
A_i &= \varepsilon_{ijk}\, x_j\, \mathcal{A}_k \nonumber \\
A_r &= -x_i\, \mathcal{A}_i \label{transversal-gauge}%
\end{align}
where $\mathcal{A}_i$ are the Cartesian components of the standard
vector potential.

These fields transform, under gauge transformations, as follows:
\begin{align}
A_i \; &\to \; g^{-1} A_i\, g - g^{-1} [L_i, g] \nonumber\\
A_r \; &\to \; g^{-1} A_r\, g - g^{-1} [P, g]
\end{align}
with
\begin{equation}
P=i\, r\, \partial_r \label{defP}
\end{equation}
Again, we are going to promote to noncommutative space the
transversal and radial fields $A_i$ and $A_r$, and not the
standard Cartesian gauge field $\mathcal{A}_i$.

We want to stress again that the map (\ref{transversal-gauge})
between spherical and Cartesian coordinates is only possible in
commutative space. In noncommutative space we are forced to work
with spherical coordinates and we cannot recover the Cartesian
coordinates. That is, in this space the fundamental fields are the
variables $A_i$, $A_r$ and not $\mathcal{A}_i$.

But now we have a problem trying to impose the constraint
(\ref{cons0}). Clearly, the constraint is not invariant under
gauge transformations and thus not well defined for gauge fields.
In order to define a gauge invariant transversal constraint we
introduce the gauge covariant distance $X_i$,
\begin{equation}
X_i = x_i -  \theta \, r\, A_i = \theta \, r\, \left(L_i - A_i
\right)
\label{bigX}
\end{equation}
As its names suggests, this quantity transforms under gauge
transformations as
\begin{equation}
X_i \; \to \; g^{-1} X_i g
\end{equation}
So the correct gauge invariant constraint is given
by\footnote{This equation was first proposed in
\cite{Karabali:2001te}. See also \cite{Nair:2001kr},
\cite{Morariu:2002tx}, \cite{Abe:2002in}.}
\begin{equation}
X_i X_i = x_i x_i = r^2
\end{equation}
This can be written as
\begin{equation}
\{ x_i, A_i \} = \theta \, r \, A_i A_i
\end{equation}
which in the limit $\theta \to 0$ coincides with (\ref{cons0}). It
is useful at this point to introduce the transverse covariant
derivative operator
\begin{equation}
D_i  = L_i - A_i \label{covder}
\end{equation}
so $X_i = \theta r D_i$ and the constraint can be written as
\begin{equation}
D_i D_i = L_i L_i = \kappa \label{consD}
\end{equation}
(we have defined $\kappa=l(l+1)=1/\theta^2$),  or
\begin{equation}
\{ L_i, A_i \} -  A_i A_i =0\label{consA}
\end{equation}

The field strength $F_{ij}$ and $F_{ir}$ are defined in analogy
with the commutative case
\begin{align}
F_{ij} &= -i\left( [L_i,A_j] - [L_j,A_i] - [A_i,A_j] -
i \varepsilon_{ijk} A_k \right) \nonumber\\
&= i\left( [D_i,D_j] - i \varepsilon_{ijk} D_k \right)
\end{align}
\begin{align}
F_{ir} &= -i\left( [L_i,A_r] - [P,A_i] - [A_i,A_r]\right)  \nonumber\\
&= i [D_i,D_r]
\end{align}
where $D_r$ is the radial covariant derivative
\begin{equation}
D_r = P- A_r
\end{equation}
For convenience we will work in the gauge $A_r=0$ so $D_r=P$.

As usual, the field $F$ is gauge covariant
\begin{align}
F_{ij} \; &\to \; g^{-1} F_{ij} g \nonumber\\
F_{ir} \; &\to \; g^{-1} F_{ir} g
\end{align}
and satisfy the transversality conditions
\begin{equation}
\{ D_i, F_{ij} \} = \{ D_i, F_{ir} \} =  0
\end{equation}
%


\section{Yang-Mills Higgs theory}

\subsection*{The action}
To write an action in this geometry, we simply write the action in
the commutative-space case in terms of transversal and radial
fields using the definition (\ref{transversal-gauge}) and then
promote the fields to noncommutative space, respecting gauge
invariance when needed. For a Yang-Mills and Higgs actions we
have\footnote{The integration is defined as $\int dx^3 = \frac{4
\pi}{2 l +1} \text{tr}\, \int r^2 dr  $, where the trace is taken
over the angular momentum representation indices.}
\begin{align}
S_{YM} &=\frac{1}{2} \; \int dx^3 \, \frac{1}{r^4} \,
{\rm tr} \left( F_{ij} F_{ij} + 2 F_{i r} F_{i r} \right) \nonumber \\
S_{Higgs}&= - \int dx^3 \left( \frac{1}{r^2} \,{\rm tr}
\left([D_i, \phi][D_i,\phi] + [D_r, \phi][D_r,\phi]\right) +
V[\phi] \right) \label{action}
\end{align}
That is, in commutative space $S_{YM}$ and $S_{Higgs}$ are the
usual Yang-Mills and Higgs actions written in term of the
transversal and radial fields $A_i$, $A_r$. Using
eqs.(\ref{transversal-gauge}) we can recover the standard form of
the actions in terms of the standard gauge potential
$\mathcal{A}_i$. However eqs.(\ref{transversal-gauge}) are not
valid in noncommutative space and expressions \eqref{action} have
to be taken as the defining actions in this geometry.


\subsection*{Equations of motion}
From actions (\ref{action}) we get the Euler-Lagrange equations of
motion
\begin{equation}
[D_r, F_{i r}] - i F_{ir} - [D_j,F_{ji}]- \frac{i}{2}
\varepsilon_{ijk} F_{jk} - i r^2 [[D_i,\phi],\phi] = \{ \mu, D_i\}
\label{eqmot1}
\end{equation}
with $\mu$ a Lagrange multiplier enforcing the constraint
(\ref{consD}). The r.h.s cancels the longitudinal part of the
l.h.s so the resulting equation is transversal.

The remaining equations of motion are
\begin{align}
&[D_i,F_{ir}] + i r^2 [[D_r,\phi],\phi] = 0 \label{eqmot2}\\
&[D_i,[D_i,\phi]] + [D_r,[D_r,\phi]] = r^2 \frac{\delta V}{\delta
\phi} \label{eqmot3}
\end{align}
We will concentrate in the case $V\equiv 0$.

To eliminate the Lagrange multiplier we note that given an
arbitrary vector $V_i$ we can write its transversal part as
\begin{equation}
 V^T_i = V_i - \frac{1}{2} \{\mu, D_i \}
 \label{trans1}
\end{equation}
for some function $\mu$. Now imposing on $V^T$ the transversality
condition for,  $\{ V^T_i, D_i\}=0$ we find the following equation
for $\mu$:
\begin{equation}
 \kappa \, \mu + D_i \mu D_i = \{ V_i, D_i \}
\label{trans2}
\end{equation}
The transversal part is obtained inserting the solution of eq.
(\ref{trans2}) in eq. (\ref{trans1}).

Before ending this section we have to mention possible BPS
equations of motion. In commutative space, in terms of the radial
and transversal fields, the BPS equations read
\begin{align}
D_r \phi& = \mp \frac{1}{2 r^2} \varepsilon_{i j k} x^i F_{j k}
\label{bps-1}\\
D_i \phi & = \pm \frac{1}{r^2} \varepsilon_{i j k} x^j F_{k r}
\label{bps-2}
\end{align}
However we have been unable to construct a noncommutative version
of them. The obvious modifications, replacing the coordinate $x^i$
by the covariant coordinate operator $X^i$ and the product of
$x^i$ and $F_{ab}$ by the Moyal anticommutator $\{ X^i, F_{ab}
\}$, does not work. For example after this replacement equation
(\ref{bps-2}) reads
\begin{equation}
[D_i, \phi] = \pm \frac{1}{2 r^2} \varepsilon_{i j k} \{X^j, F_{k
r} \}
\end{equation}
But while the l.h.s is transversal with respect to $X^i$, the
r.h.s is not. Even projecting the r.h.s over the transverse
components does not reproduce the equations of motion.




\section{Monopole solutions}

\subsection*{Spherically symmetric ansatz}
The most general spherically symmetric ansatz can be written using
the operators
\begin{equation}
V^{(0)}_i = L_i \; , \;\; V^{(1)}_i = \sigma^i \; , \;\; V^{(2)}_i =\{\alpha, L_i\} \; ,
\;\; V^{(3)}_i = [\alpha, L_i] \label{operators}
\end{equation}
where $\sigma^i$ are the Pauli matrices and
\begin{equation}
\alpha = \sum_i^3 \;\sigma_i L_i
\end{equation}

Although (\ref{operators}) is the most general set of rotationally
covariant operators, it can be shown that the set remains
consistent if we drop $V^{(3)}_i$. That is, when we expand the
fields in the basis $\{V^{(0)}_i, V^{(1)}_i, V^{(2)}_i\}$ the
equations of motions does not have components in the direction
$V^{(3)}_i$. So, from now on we will work with the basis
$\{V^{(0)}_i, V^{(1)}_i, V^{(2)}_i\}$.

Then we expand
\begin{equation}
D_i= \sum_{a=0}^2 \; v_{a} V^{(a)}_i
\end{equation}
with $v_a$ arbitrary functions of the radial coordinate $r$,
$v_a\equiv v_a(r)$.


The constraint (\ref{consD}) implies the following two equations
for the coefficients $v_0$, $v_1$ and $v_2$
\begin{align}
&\kappa \,{v_{0}}^2 + 3\,{v_{1}}^2 + 4\,\kappa \,v_{1}\,v_{2} +
  2\,\kappa \,\left( 2\,\kappa - 1 \right) \,{v_{2}}^2 = \kappa \nonumber \\
&  \left( 4\,\kappa - 3 \right) \,{v_{2}}^2 - 2\,v_{0}\,\left(
v_{1} - \left( 2\,\kappa - 1 \right) \,v_{2} \right)  = 0
\label{spher-consD-2}
\end{align}
That is, the field $D_i$ depends only on one function.


We have for the field strength
\begin{align}
F_{ij}= \varepsilon_{i j a} &\left\{ \left( v_{0} - {v_{0}}^2 +
     \left( 3 - 4\,\kappa \right) \,{v_{2}}^2 \right)
     \,V^{(0)}_a + \right. \nonumber\\
& \left( -2\,{v_{1}}^2 + 2\,\kappa \,v_{0}\,v_{2} -
     4\,\kappa \,{v_{2}}^2 + v_{1}\,\left( 1 - 4\, \kappa \,v_{2} \right)
     \right) \,V^{(1)}_a + \nonumber\\
& \left.
  \left( v_{2} - 3\,v_{0}\,v_{2} + 2\,v_{1}\,v_{2} +
     5\,{v_{2}}^2 \right) \,V^{(2)}_a \right\}
\end{align}
and
\begin{align}
F_{kr}= r \, \left( v_0'\, V^{(0)}_k + v_1'\, V^{(1)}_k + v_2'\,
V^{(2)}_k \right)
\end{align}

To write the equation of motion \eqref{eqmot1} we have first to
solve equation \eqref{trans2} to find the Lagrange multiplier
$\mu$ that projects the solution onto the tangential space.
Spherical symmetry imposes that $\mu$ has the form
\begin{equation}
\mu = \mu_0 + \alpha \, \mu_1
\end{equation}
so equation (\ref{trans2}) leads to algebraic equations for the
coefficients $\mu_0$ and $\mu_1$. However we will see later that
we can solve explicitly the constraint and thus work with the
physical, unconstrained degrees of freedom, making unnecessary the
Lagrange multiplier.

The Higgs field can be expanded as
\begin{equation}
\phi= \phi_0(r) + \phi_1(r) \, \alpha
\end{equation}
and the covariant derivatives takes the form
\begin{align}
[D_i,\phi] &=  -\phi_1 \, (v_0 - 2\, v_1 - v_2)\,  V^{(3)}_i
\nonumber \\
[D_r,\phi] &=  i\,r\, \left( \phi_0'(r) + \phi_1'(r) \, \alpha
\right)
\end{align}
It can be checked that the equation of motion (\ref{eqmot2}) is
trivially satisfied.

Finally, the last equations, (\ref{eqmot3}) take the form
\begin{align}
r \frac{d^2}{dr^2} \left( r \phi_0 \right) &= 0 \;\;\;\;\; \to
\;\;\;\; \phi_0 = \frac{c_1}{r} + c_0 \nonumber \\
r \frac{d^2}{dr^2} \left( r \phi_1 \right) &= 2 \phi_1 \left(
v_0-2 v_1- v_2 \right)^2 \label{phieqs}
\end{align}
Notice that $\phi_0$ is decoupled from the other fields and in
fact it is an irrelevant constant.

In these variables the Hamiltonian takes the form
\begin{align}
H =& {8\,\pi}\,\int d r \, \left\{ r^2\,
      \left( {\phi_0'}^2 +
        {\phi_1'}^2\,\kappa \right)  +
     2\,{\phi_1}^2\,\kappa\,
      {\left( -v_0 + 2\,v_1 + v_2 \right) }^2
     \right)  + \nonumber \\
&  \frac{1}{{r^2}} \, \left( 3\,{\left( 1 - 2\,v_1 \right)
}^2\,{v_1}^2 +
       \kappa\,\left( -2\,{v_0}^3 +
       {v_0}^2\,\left( 1 + 8\,\left( 4\,\kappa - 3 \right) \,{v_2}^2 \right)
       - \right. \right. \nonumber \\
&  \left. \left.   2\,\left( 4\,\kappa - 3\right) \,v_0\,{v_2}^2\,
           \left( 3 + 4\,v_1 + 10\,v_2 \right)  +
          v_2\,\left( 32\,{v_1}^3 + 8\,{v_1}^2\,\left(4\,\kappa\,v_2 - 3 \right)
          + \right. \right. \right. \nonumber \\
&  \left. \left. \left.  v_2\,\left( -2 + 4\,\kappa + 4\,\left(
6\,\kappa - 5 \right) \,v_2 +
                \left( -41 + 4\,\kappa\,\left( 11 + 4\,\kappa \right)  \right) \,{v_2}^2 \right)
                + \right. \right. \right.  \nonumber \\
&  \left. \left. \left.  4\,v_1\,\left( 1 + v_2\,\left( -3 +
2\,\left( 8\,\kappa - 5 \right) \,v_2 \right) \right) \right)
\right) +   {v_0}^4
       \right)  + \left( 3\,{{v_1'}}^2 + \kappa\,\left( {{v_0'}}^2
       + \right. \right. \nonumber \\
& \left. \left.  2\,{v_2'}\,
         \left( 2\,{v_1'} + \left( -1 + 2\,\kappa \right) \,{v_2'} \right)  \right)
         \right\}
\end{align}
%




\subsection*{Small $\theta$ expansion}

Let us study first the small $\theta$ expansion of the monopole
equations. First we note that the operator $V^{(2)}_i$ is already
of order $1/\theta^2$:
\begin{equation}
V^{(2)}_i =\{\alpha, L_i\} =\frac{1}{\theta^2} \, \{ {\hat X}
\cdot {\vec \sigma}, {\hat X}^i \}
\end{equation}
so the coefficient $v_2$ is of order $\theta^2$. (Since the
covariant derivative operator starts with $L_i$ at zero-order the
coefficient $v_0$ is order zero).

To compare with the usual 't Hofft\~-Polyakov\~-Julia\~-Zee
\-Prasad \-Sommer\-field solutions we write
\begin{align}
&v_0 - 1 = \theta^2 \, v_0^{(2)} + \theta^4 \, v_0^{(4)} + \cdots \nonumber \\
&v_1= -\frac{k-1}{2} - \theta^2 \frac{k_1}{2} + \cdots\nonumber\\
&v_2=  v_2^{(0)}\theta^2 + \theta^4 \, v_2^{(2)} + \cdots \nonumber\\
&\phi_1 = \frac{\theta}{2r} \left(h + \theta^2 h_1 +\cdots \right.
\end{align}
The constraints (\ref{spher-consD-2}) can be solved perturbatively
in $\theta$ and we can write the coefficients of $v_0$
($v_0^{(2)}, v_0^{(4)}, \cdots$) and $v_2$ ($v_2^{(0)}, v_2^{(2)},
\cdots$) as functions of the coefficients of $v_1$ ($k, k_1,
\cdots$). We have
\begin{align}
&v_0 - 1= - \frac{\theta^2}{4} {\left(k-1 \right) }^2
    - \frac{\theta^4}{32}\, \left(k-1 \right)\,
     \left( 5 + k - 7\,k^2 + k^3 + 16\,k_1 \right)  + \cdots \nonumber \\
&v_1 = \frac{1 - k}{2} + \frac{\theta^2}{8}\,
     \left(k^2-1 - 4\,k_1 \right)  \cdots
     \nonumber\\
&v_2 =\theta^2\, \frac{k-1}{4} + \theta^4 \, \frac{1}{4}\, k_1 +
\cdots
     \nonumber\\
\end{align}

At leading order we recover the standard monopole equations
\begin{align}
& r^2\, k''(r) = k(r)\left( k(r)^2-1 + h^2(r)\right)  \nonumber
\\
&r^2 {h}''(r) = 2 k(r)\, h(r)
\end{align}
with the well-known solutions \cite{Prasad:1975kr}
\begin{align}
k(r) &= \frac{r}{\sinh (r)} \nonumber\\
h(r) &= r\, \coth (r) -1
\end{align}

 The next order equations read
\begin{align}
r^2\,&{k_{1}}''(r)\, +  \left( 1 - {{h}(r)}^2 - 3\,{k(r)}^2
\right) \,{k_{1}}(r) = \frac{1}{4}\left( -1 +
8\,{h}(r)\,{h_1}(r)\, k(r) +   \right. \nonumber \\
& 3\,{k(r)}^2 + 7\,{k(r)}^3 - 4\,{k(r)}^4 -
 2\,{k(r)}^5 + {{h}(r)}^2\,
\left( 1 + k(r) -  \right. \nonumber\\
& \left. \left. 2\,{k(r)}^3 \right)  +  4\,r^2\,{k'(r)}^2 + k(r)\,
   \left( -3 - 2\,r^2\,{k'(r)}^2 + 2\,r^2\,k''(r) \right) \right)
\end{align}
\begin{align}
r^2\,{h_1}''(r) - 2\,{h_1}(r)\,{k(r)}^2 =
   {h}(r)\,k(r)\,
   \left( 1 + k(r) - 2\,{k(r)}^2 + 4\,{k_{1}}(r) \right)
\end{align}
and can be solved numerically.




\subsection*{Solving the constraint}

Instead of working with the ``linear" variables $v_0, v_1, v_2$
and the constraints  (\ref{spher-consD-2}) we can try to
reparametrize the fields and solve the constraint explicitly. Then
the resulting fields are the physical degrees of freedom and the
constraint is automatically incorporated in the equations of
motion.

In fact, we can see that the replacement
\begin{align}
v_0 &\to  (z_0 - z_1)/{\sqrt{2}} \nonumber \\
v_1 &\to  (z_0 + z_1)/{\sqrt{2}} - \frac{2\kappa -1}
{\sqrt{4\kappa -3}}\, z_2 \nonumber \\
v_2 &\to {z_2}/{\sqrt{4\kappa -3}}
\end{align}
diagonalizes the second equation (\ref{spher-consD-2})
\begin{equation}
z_0^2 - z_1^2 - z_2^2=0
\end{equation}
which is straightforwardly solved in term of two functions $\rho$
and $u$ (both are functions of $r$)
\begin{align}
z_0 & = \rho \nonumber \\
z_1 & = -\rho\, c(u) \nonumber \\
z_2 & = -\rho \, u
\end{align}
with
\begin{equation}
c(u)=\sqrt{1-u^2} \;\;\;\;\;\;\;\;\;\; \text{and}
\;\;\;\;\;\;\;\;\; -1\leq u \leq1 \label{def-c}
\end{equation}
(we chose the branch solution that matches the standard $\theta
\to 0$ limit). Replacing this solution into the first of equations
(\ref{spher-consD-2}) we get a quadratic equation for $\rho$ that
can be easily solved. Finally we have a parametrization that
solves the constraint
\begin{align}
v_0 & = \frac{\sqrt{\kappa }}{\sqrt{2\,d(u)}}\,\left( 1 + c(u)\right) \nonumber \\
v_1 & = \frac{\sqrt{\kappa }}{2\,\sqrt{d(u)}\,\sqrt{4\kappa -3}}
\left(2 \,
u\,(2\kappa -1) + \sqrt{8\kappa -6}\,\left(1-c(u)\right)\right) \nonumber\\
v_2 & = - \frac{\sqrt{\kappa }}{\sqrt{d(u)}\,\sqrt{4\kappa -3}}\,
u
\end{align}
where
\begin{align}
d(u)= 3 + \kappa + \frac{1}{2}\,u^2\,\left(3\,\kappa - 5\right) +
\left( \kappa -3\right)\,c(u) + \sqrt{8 \kappa
-6}\,\left(1+c(u)\right)\,u \label{def-d}
\end{align}
That is, we have parametrized the gauge fields in terms of only
one function $u$, which together with the Higgs field $\phi_1$ are
the only nontrivial degrees of freedom. The next step is to write
the equations of motion in terms of them. Actually, though we have
reduced significantly the number of degrees of freedom, the
equations of motion are very complicated in terms of these fields.
We show the complete expression of the equations of motions in the
appendix.

In this variables the small $\theta$ limit can be recovered
through the identification
\begin{equation}
u(r) = \frac{\theta}{\sqrt{2}} \left( 1 - k(r)\right) +
O(\theta^3)
\end{equation}
%




\subsection*{Boundary conditions}

In order to get non-singular, finite energy solutions we have to
impose appropriate boundary conditions. At the origin we have the
usual conditions
\begin{align}
&u(0)=0 \nonumber \\
&\phi_1(0)=0
\end{align}
At $r \to \infty$ the situation is different from the commutative
case. Notice that in the presence of a potential,
\begin{equation}
V=\lambda\, \left( \phi^2 - \eta^2 \right)^2
\end{equation}
the Higgs field tends asymptotically to a minimum of the
potential. That is, asymptotically, $\phi_0$ and $\phi_1$ are
minima of
\begin{equation}
V = \lambda \, \left( \frac{1}{16} - \frac{1}{2}\, \phi_0^2 +
\phi_0^4 - \frac{\kappa}{2}\, \phi_1^2 +6\, \kappa\, \phi_0^2\,
\phi_1^2 - 4\, \kappa\, \phi_0\, \phi_1^3 + \kappa\, \left( \kappa
+ 1\right) \phi_1^4\right)
\end{equation}

\begin{equation}
V=1 - 2\,{\phi_{0}}^2 + {\phi_{0}}^4 -
  2\,{\phi_{1}}^2\,\kappa + 6\,{\phi_{0}}^2\,{\phi_{1}}^2\,\kappa -
  4\,\phi_{0}\,{\phi_{1}}^3\,\kappa + {\phi_{1}}^4\,\kappa +
  {\phi_{1}}^4\,\kappa ^2
\end{equation}
(we have rescaled the fields so $\eta=1/2$, consistent with the
small $\theta$ expansion solution). Besides the trivial solution
$\phi_0=1, \; \phi_1=0$ we have the solutions
\begin{align}
&\phi_0 = \frac{1}{4}\left(1 + \frac{1}{{\sqrt{1 + 4\,\kappa
}}}\right) \, ,
&\phi_1 = \frac{1}{2{\sqrt{1 + 4\,\kappa }}} \\
&\phi_0 = \frac{1}{2{\sqrt{1 + 4\,\kappa }}} \; , & \phi_1 =
\frac{1}{{\sqrt{1 + 4\,\kappa }}}
\end{align}
(these correspond to absolute minima of the potential; there are
other local minima but those will give infinite energy when
integrated over the whole space).

The first of these equations gives a nontrivial $U(1)$
contribution in the $\theta \to 0$ limit so we discard it. The
second one gives the correct small $\theta$ behavior so we take it
as the asymptotic boundary condition:
\begin{align}
%
\lim_{r\to \infty} \phi_1(r) = \frac{1}{{\sqrt{1 + 4\,\kappa }}}
\end{align}
Of course, this is valid in the presence of a potential. For
vanishing coupling constant, as it happens in commutative space,
we can rescale the Higgs fields arbitrarily by rescaling
appropriately the radial variable $r$.

For the gauge field we impose, as usual, that at infinity the
Higgs kinetic term vanishes. This gives the behavior
\begin{equation}
\lim_{r\to \infty} u(r) = \frac{4\,{\sqrt{2}}} {{\sqrt{4\,\kappa -
3}} + 3\,{\sqrt{4\,\kappa +1}}}
\end{equation}
%



\subsection*{Numerical solutions}

\begin{figure}[h]
\begin{center}
\includegraphics[angle=0, width=13cm]{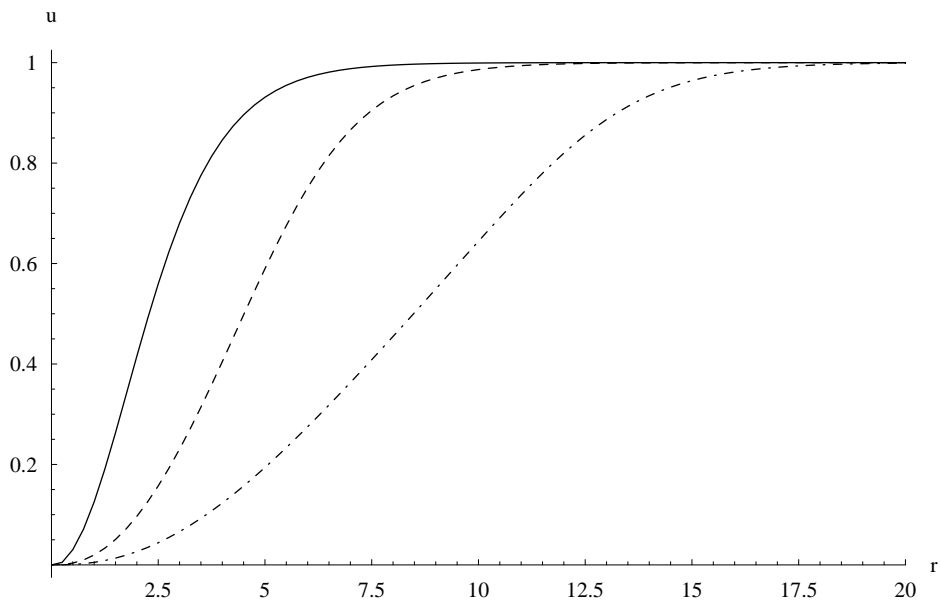}
\end{center}
\smallskip
\caption{The field $u(r)$ (normalized to $1$ at infinity) for
different values of $\kappa$. The solid line is for $\kappa=10$
(indistinguishable from the standard BPS solution), the dashed
line is for $\kappa=0.9$ and the dot-dash line is for
$\kappa=0.76$.} \label{fig-u}
\end{figure}

\begin{figure}[h]
\begin{center}
\includegraphics[angle=0, width=13cm]{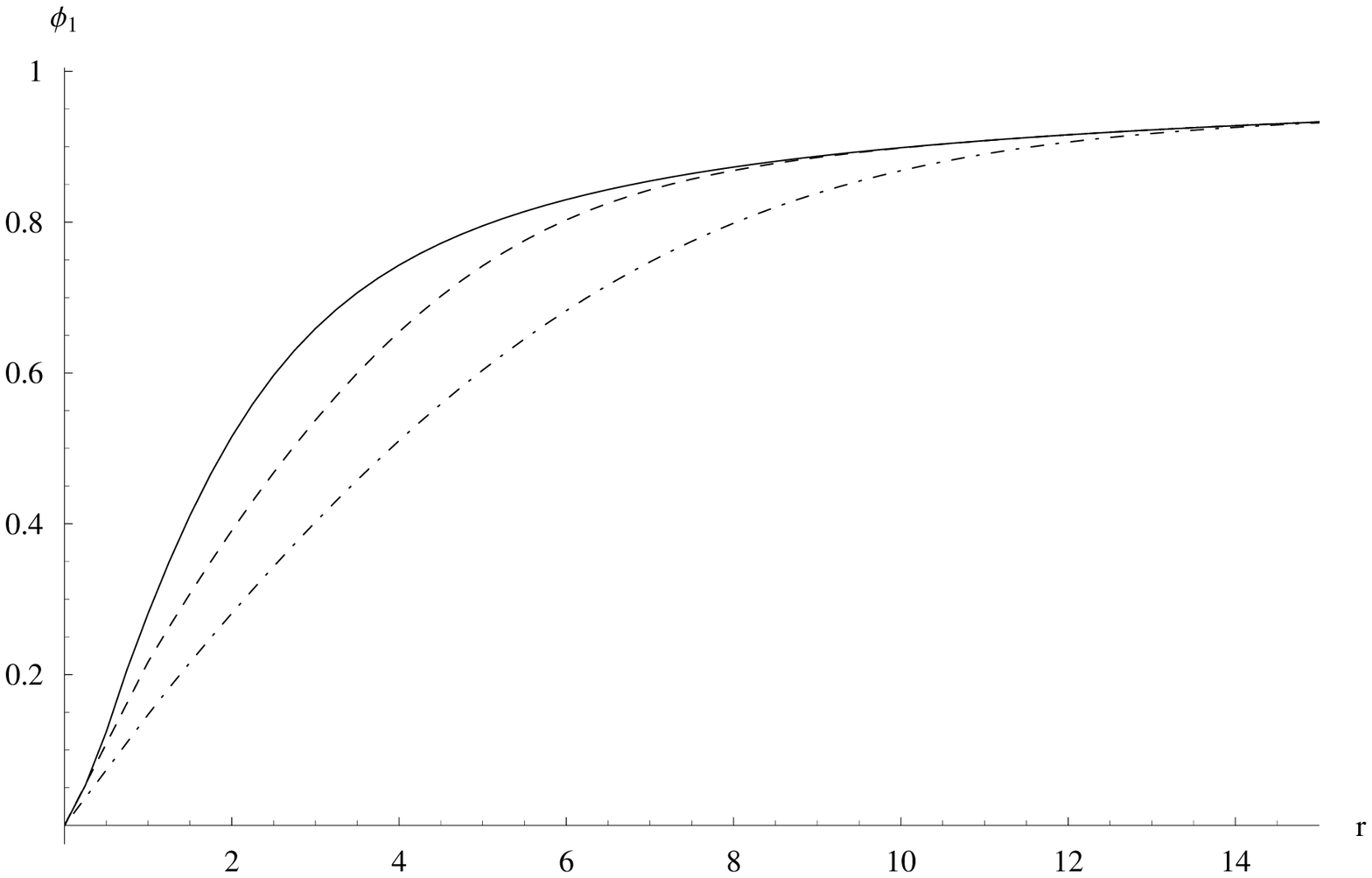}
\end{center}
\smallskip
\caption{The field $\phi_1(r)$ (normalized to $1$ at infinity) for
different values of $\kappa$. The solid line is for $\kappa=10$
(indistinguishable from the standard BPS solution), the dashed
line is for $\kappa=0.9$ and the dot-dash line is for
$\kappa=0.76$.} \label{fig-phi}
\end{figure}

We solved numerically the equations of motion for different values
of $\kappa =1/\theta^2=l(l+1)$. We found solutions for essentially
any value of $\kappa $ allowed ($\kappa >3/4$). As expected, for
large $\kappa $ (small $\theta$ ) the solution tends to the
Prasad-Sommerfield configurations. Indeed, even for $l=1$, the
profile of the solutions are very similar to the P-S solutions. In
order to see the departure of the P-S solutions we considered
continuos values of $\kappa $ (which correspond to infinite
dimensional representation of the non-commutative algebra). It is
remarkable that the Higgs field solution is not very sensitive to
$\kappa $, even for extreme values ($\kappa \sim 3/4$). On the
other hand, the gauge field in very sensitive to $\kappa $. We
show in figures \ref{fig-u} and \ref{fig-phi} the solutions for
the fields $u$ and $\phi_1$ respectively, for various values of
$\theta$.

\begin{figure}[h]
\begin{center}
\includegraphics[angle=0, width=13cm]{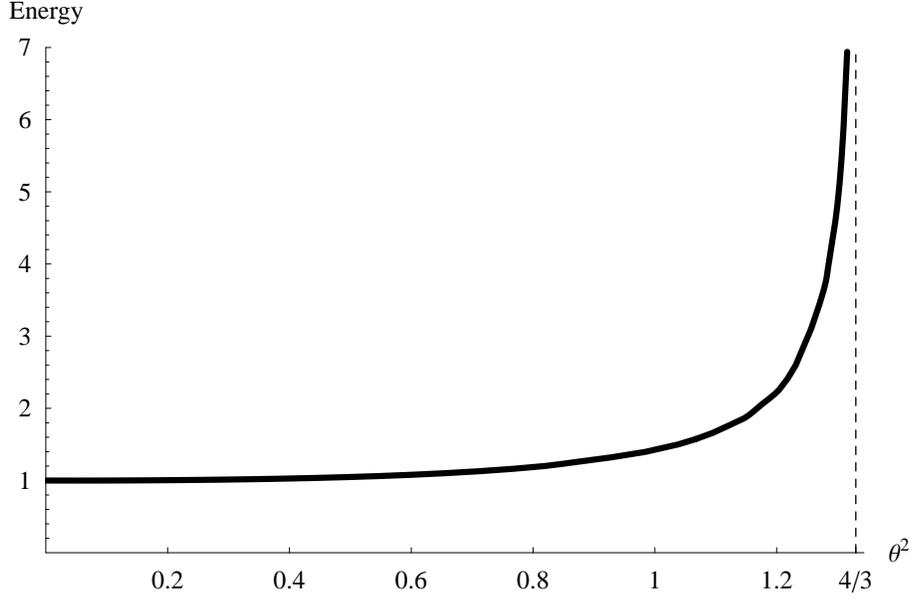}
\end{center}
\smallskip
\caption{Energy, in units of $e^2/4 \pi$, of the monopole as a
function of $\theta^2$. The energy tends to $1$ when $\theta^2 \to
0$ (commutative BPS solution) and diverges when $\theta^2\to
4/3$.} \label{fig-energy}
\end{figure}

We also studied the energy of the monopole solutions as a function
of $\theta$. For small vales of $\theta$, the energy, in units of
$e^2/4 \pi$, tends to $1$ as expected (BPS bound). As $\theta$
increases, the energy also increases and diverges as $\theta^2$
approaches to $4/3$. A plot of the energy as a function of
$\theta$ is shown in figure \ref{fig-energy}. This behavior is
another hint that for $\theta$ different from zero, the solutions
obtained are not self-dual, since in that case we expect the
energy of the configuration to be equal to some topological number
(independent of $\theta$). This situation can be contrasted with
the case of self-dual vortex solutions in NC space. In the later,
while the profile of the solutions are dependent of the
noncommutative parameter, the energy is $\theta$-independent and
in particular equal to $1$ (in appropriate units), the Bogomolny
energy bound \cite{Lozano:2000qf}.

\section{Summary and conclusions}

Previous analysis of monopole configuration in noncommutative
space were done using the standard non-commutative relations
\begin{equation}
[x_i,x_j] = i \,\theta_{ij} \;\;\;\;\;\; \theta_{ij} :
\text{constant} \label{summ-1}
\end{equation}
Although this algebra is invariant under space translations, as is
immediate from the definition, commutation relations
\eqref{summ-1} are not invariant under space rotations. In
particular we cannot benefit from the simplifications, in
structure and in number of degrees of freedom, that a spherically
symmetric ansatz produce.

In contrast to relations \eqref{summ-1}, we can construct a
different noncommutative algebra which is manifestly rotationally
invariant
\begin{equation}
[x_i, x_j] = i \theta\, r\, \varepsilon_{i j k}\, x_k
\label{summ-2}
\end{equation}
but at the expense of loosing translational invariance. In fact,
the algebra \eqref{summ-2} is incompatible with a momentum
operator $P_i$ generating infinitesimal translations. However,
this is not an impediment to construct a field theory in this
geometry. A representation of this algebra can be constructed by
identifying $x_i= \theta r L_i$. In this representation the value
$\theta$ labels the representation through the relation
$1/\theta^2 = {\vec L}^2$. So, although $\theta$ can take any
positive value, for the special case $1/\theta^2 = l(l+1)\, , \;
l\in \mathbb{N}$, we have finite-dimensional representations
(notice however that the radial variable $r$ takes continuous
values).

In commutative space, a Poincar\'e invariant Lagrangian can be
written in terms of momentum operators $P_i$,where translational
invariance is manifest, or in terms of angular momentum operators
$L_i$ (together with a radial scaling operator $P$), where
rotational invariance is obvious. To construct a NCFT with the
algebra \eqref{summ-1} one choose the former and promote the
variables (with some prescribed order) to noncommutative
operators. Analogously, to construct a NCFT with the algebra
\eqref{summ-2}, we can choose the later and again, promote the
variables to noncommutative operators.

In particular we constructed a Yang-Mills Higgs Hamiltonian in
this space, and also derived the equations of motions. A puzzling
aspect is that we were unable to derive first order (BPS)
equations of motion. Though we do not have a rigorous proof of
this statement, there are several hints that suggest this
property. Since the theory is manifestly invariant under rotations
we tried a spherically symmetric ansatz, which is nothing but a
noncommutative extension of the 't Hooft monopole ansatz. Then, as
it happens in the commutative case, the number of degrees of
freedom is reduced to just two, one for the gauge field and other
for the Higgs field. The final equations of motion are very
complicated in form but not difficult to solve numerically.
Moreover, we showed that in the limit $\theta \to 0$ the equations
of motion (and in fact the whole Hamiltonian), reduces to the
standard commutative Yang-Mills-Higgs theory, allowing then a
perturbative solution in the noncommutative parameter. Another
characteristic of this theory is that it blows up at
$\theta^2=4/3$, which incidentally is the maximum value of
$\theta$ for which there is a finite-dimensional representation of
the algebra.

We solved numerically the Euler equations of motion for different
values of $\theta\,$. As expected for small values of $\theta\,$
the solutions is indistinguishable from the exact
Prasad-Sommerfield solution. As we increase the value of
$\theta\,$ the profile of the solution depart from the P-S
solutions, and also the energy increases. In particular the energy
diverges as $\theta\,$ approaches to its maximum value
$\theta^2=4/3$.




\begin{acknowledgments}
We are grateful with Diego Correa and Gustavo Lozano for helpful
comments. We would like to thank Fidel Schaposnik for valuable
discussions and critical readings of the manuscript.
\end{acknowledgments}



\appendix*
\section*{Appendix}

In this appendix we present the Euler-Lagrange equations of motion
in terms of the unconstrained variable $u$ and the Higgs field
$\phi_1$. The equations read:

\begin{align*}
&u''(r) + \left( \left( 2\,r^2\,\kappa\,{c( \,u)}^4\,
        {{d}( \,u)}^2\,
        {{\phi_1}(r)}^2\,
        \left( {\sqrt{8\,\kappa -6}} - 3\,{\sqrt{8\,\kappa -6}}\,
           c( \,u) + \right. \right. \right. \nonumber \\
& \left. \left. \left. 2\, \,\left( 4\,\kappa - 3 \right) \,
   u \right) \, \left( 6\,\left( 4\,\kappa - 3 \right) \,
   {c( \,u)}^3 + {c( \,u)}^2\, \left( 24 - 32\,\kappa +
   \,\left( 21 - 17\,\kappa \right) \times\right. \right. \right.\right. \nonumber \\
& \left. \left. \left. \left.    {\sqrt{8\,\kappa -6}}\,u \right)
   + \,u\,
           \left( \left( 3 - \kappa \right) \,
              {\sqrt{8\,\kappa -6}} -
             6\,{\sqrt{8\,\kappa -6}}\,
              {d}( \,u) -
              \right. \right. \right. \right. \nonumber \\
& \left. \left. \left. \left.  2\, \,
              \left( 12 - 19\,\kappa + 4\,\kappa^2 \right) \,u +
              2\,{\sqrt{2}}\, {\left( 4\,\kappa -
3 \right) }^{\frac{3}{2}}\,
              {u}^2 \right)  + 2\,c( \,u)\,
           \left( 4\,\kappa - 3 +
             \right. \right. \right. \right. \nonumber \\
& \left. \left. \left. \left.
              \left( 6 - 8\,\kappa \right) \,
              {d}( \,u) +  \left(7\,\kappa - 10\right) \,
              {\sqrt{8\,\kappa -6}}\,u +
             \left( 24 - 41\,\kappa + 12\,\kappa^2 \right) \,
              {u}^2 \right)  \right) \right) \times \right. \nonumber \\
& \left.  \left({4\,\kappa - 3}\right)^{-1}
              + {c( \,u)}^4\, {d}( \,u)\,
      \left( {\sqrt{2}}\,\kappa\,
         \left( -\left( {\sqrt{2}}\,\kappa\,
              {\left( 1 + c( \,u)
                  \right) }^3 \right)  +  \right. \right. \right. \nonumber \\
&  3\,{\sqrt{\kappa}}\,
            {\left( 1 + c( \,u)
                \right) }^2\,
            {\sqrt{{d}( \,
                 u)}} - {\sqrt{2}}\,\left( 1 +
              c( \,u) \right) \, \left( {d}( \,  u) +
              8\, \kappa\,{u}^2 \right)
            +  \nonumber \\
&  \left( 2\, {\sqrt{\kappa}}\,
              {u}^2\,
              \left( -2\,{\sqrt{2}}\,{\sqrt{\kappa}}\,
                 {\sqrt{4\,\kappa - 3}}\,
                 c( \,u) + 3\,{\sqrt{4\,\kappa - 3}}\,
{\sqrt{{d}(\,u)}} +  \right. \right. \nonumber \\
& \left. \left. \left. \left. 2\,{\sqrt{\kappa}}\,
                 \left( {\sqrt{8\,\kappa -6}} +
                 \, \left( -7 + 4\,\kappa \right) \,u
                   \right)  \right) \right) \left(4\,\kappa -
                   3\right)^{-1/2}
           \right) \times \left( -2\, \,  {d}( \,u)\,
          u + \right. \right. \nonumber \\
&   \left( 1 +  c( \,u) \right) \,
          \left( {\sqrt{8 \,\kappa -6}}\, {c( \,u)}^2 -
          c( \,u)\, \left( {\sqrt{8\,\kappa -6}} +
         \, \left(3\,\kappa - 5 \right) \,u \right)
               + \right.  \nonumber \\
&   \left. \left. \left. \left. \left.  u\, \left( \kappa - 3 -
                 \,{\sqrt{8\,\kappa -6}}\,
                  u \right)  \right)  \right)  +
        \left(2\, \,\kappa\,u\,
           \left( -2\,{\sqrt{4\,\kappa - 3}}\,
              {c( \,u)}^3 + \right. \right. \right. \right.\right. \nonumber \\
&   \left. \left. \left. \left.     {c( \,u)}^2\,
              \left( 4\,{\sqrt{4\,\kappa - 3}}
             +  {\sqrt{2}}\, \, \left( 3\,\kappa - 5 \right) \,u \right)
              + u\, \left( {\sqrt{2}}\,\left( \kappa - 3 \right)  +
                2\,{\sqrt{2}}\, {d}( \,u) -
                \right. \right. \right. \right. \right. \nonumber \\
& \left. \left. 2\, \, {\sqrt{4\,\kappa - 3}}\,u \right)  +
                     2\,c( \,u)\, \left(-{\sqrt{4\,\kappa - 3}} - 2\,{\sqrt{2}}\, \,
                 \left( \kappa - 2 \right) \,u +
                 {\sqrt{4\,\kappa - 3}}\,{u}^2 \right)
             \right) \times  \nonumber \\
& \left. \left. \left. \left( 24\,\kappa\,
              \left( 4\,\kappa - 3 \right) \,
              {c( \,u)}^2 +
             \left( 8\,\kappa - 6 \right) \,
              {d}( \,u) -
             3\,{\sqrt{\kappa}}\,
              {\sqrt{{d}( \,
                   u)}}\,
              \left( {\sqrt{2}}\,
                 \left( 4\,\kappa - 3 \right)  +
              \right. \right. \right. \right. \right. 
%
%
\end{align*}
\begin{align}
%
&        \left. 2\, \, \left( 7 - 4\,\kappa \right) \,
                 {\sqrt{4\,\kappa - 3}}\,u \right)  +
             \kappa\,u\, \left( \left( 27 - 20\,\kappa \right) \,
                 {\sqrt{8\,\kappa -6}} +
                2\, \,
                 \left( 71 - 64\,\kappa + \right. \right. \nonumber \\
&  \left. \left. 16\,\kappa^2 \right) \,
                 u \right)  + c( \,u)\,
              \left( 15\,{\sqrt{2}}\,
                 \left( 3 - 4\,\kappa \right) \,{\sqrt{\kappa\, d(u)}}\, +
                3\,\kappa\,\left( 8\,
                   \left( 4\,\kappa - 3 \right)  + \right. \right. \nonumber \\
&   \left. \left. \left. \left. \left( 29 - 12\,\kappa \right) \,
                    {\sqrt{8\,\kappa -6}}\,u \right)
                \right)  \right) \right)  \times   \left({\left( 4\,\kappa - 3
              \right) }^{\frac{3}{2}}\right)^{-1} +
        \left( \left( -6\,\left( 4\,\kappa - 3 \right) \,
              {c( \,u)}^3 + \right. \right. \nonumber \\
&       {\sqrt{4\,\kappa - 3}}\, {c( \,u)}^2\,
              \left( 12\,{\sqrt{4\,\kappa - 3}} + {\sqrt{2}}\, \,
                 \left( -21 + 17\,\kappa \right) \,u \right)
                 + \nonumber \\
&   u\, \left( 3\,\left( \kappa - 3 \right) \,
                 {\sqrt{8\,\kappa -6}} +
                6\,{\sqrt{8\,\kappa -6}}\,
                 {d}( \,u) + \ 2\, \, \left( 18 - 27\,\kappa + 4\,\kappa^2 \right) \,u
                 + \right. \nonumber \\
&       \left.    2\, \left( 3 - 4\,\kappa \right) \,
  {\sqrt{8\,\kappa -6}}\,{u}^2 \right)  +   2\,c( \,u)\,
              \left( 9 - 12\,\kappa +
                \left( 8\,\kappa - 6 \right) \,
                 {d}( \,u) - \right.  \nonumber \\
&    \left. \left. 5\, \, \left( 2\,\kappa - 3 \right) \,
                 {\sqrt{8\,\kappa -6}}\,u +  \left( -24 + 41\,\kappa - 12\,\kappa^2 \right) \,
                 {u}^2 \right)  \right) \times \nonumber \\
&      \left( 4\,{\sqrt{2}}\,\kappa\,
              {\left( 4\,\kappa - 3 \right) }^{\frac{3}{2}}\,
              {c( \,u)}^3 -  \left( 4\,\kappa - 3 \right) \,
              {d}( \,u)\,\left( {\sqrt{8\,\kappa -6}} +
                2\, \,
                 \left( 2\,\kappa - 1 \right) \,u \right)
              + \right. \nonumber \\
&    \left. 6\,{\sqrt{\kappa}}\,{\sqrt{4\,\kappa - 3}}\,
              {\sqrt{{d}( \, u)}} \times  \left( 4\,\kappa - 3 +
                \left( -2 + 3\,\kappa \right) \,
                 {\sqrt{8\,\kappa -6}}\,u +
                2\, {u}^2 \right)  - \right. \nonumber \\
&         2\,\kappa\,\left( 2\,{\sqrt{2}}\,
                 {\left( 4\,\kappa - 3 \right) }^{\frac{3}{2}}
                 + 4\, \, {\left( 3 - 4\,\kappa \right) }^2\,u +
               3\, {\sqrt{8\,\kappa -6}}\,
                 \left( 4 - 7\,\kappa + 4\,\kappa^2 \right) \,
                 {u}^2 -  \right. \nonumber \\
&    \left.            2\, \left( 4 - 43\,\kappa + 48\,\kappa^2 -
                16\,\kappa^3 \right) \,{u}^3 \right)  +
             6\,{c( \,u)}^2\,
              \left(\left( {\sqrt{\kappa}}\,
                   {\left( 4\,\kappa - 3 \right) }^{\frac{3}{2}}\,
     {\sqrt{{d}( \,u)}} \right)  - \right. \nonumber \\
&    \left.  2\,\kappa\,\left( 4\,\kappa - 3 \right) \times
                \left( {\sqrt{8\,\kappa - 6}} + 2\, \, \left( -1 + 2\,\kappa \right) \,u
                   \right)  \right)  +
             {\sqrt{4\,\kappa - 3}}\,
              c( \,u) \times \nonumber \\
&                \left( {\sqrt{2}}\,
                 \left( 4\,\kappa - 3 \right) \, {d}( \,u) - 6\,{\sqrt{\kappa}}\,
                 {\sqrt{{d}(u)}}\, \left( 8\,\kappa - 6 +
                   \left( \kappa - 2 \right) \,
                    {\sqrt{8\,\kappa -6}}\,u \right)  + \right. \nonumber \\
&             \left. \left. \left.
                6\,\kappa\,\left( {\sqrt{8}}\, \left( 4\,\kappa
                - 3 \right)  +  \left( \kappa - 1 \right) \,
                    {\sqrt{4\,\kappa - 3}}\,u +
                   {\sqrt{2}}\, \left( 4 - 21\,\kappa + 12\,\kappa^2 \right) \,
                    {u}^2 \right)  \right)  \right) \right) \times
                    \nonumber \\
&       \left. \left. \left(\left( 3 - 4\,\kappa \right)
            \right)^{-2} \right)  -  r^2\,\kappa\,
      \left( -16\,{\sqrt{8\,\kappa -6}}\,
         \left( 12 + \kappa^2 +
           \left( -12 + \kappa^2 \right) \,
            c( \,u) \right)  + \right. \right.\nonumber \\
&      16\, \,
         \left( 117 - 54\,\kappa - 8\,\kappa^2 + 2\,\kappa^3 +
           \left( -117 + 58\,\kappa - 8\,\kappa^2 + 2\,\kappa^3 \right)
              \,c( \,u) \right) \,u  + \nonumber \\
&         4\, {\sqrt{8\,\kappa -6}}\,
         \left( 339 - 58\,\kappa + 17\,\kappa^2 +
           \left( -315 + 38\,\kappa + 15\,\kappa^2 \right) \,
            c( \,u) \right) \,{u}^2 - \nonumber \\
&         4\, \left( 1101 - 884\,\kappa -  55\,\kappa^2 +
               30\,\kappa^3 -
           \left( 859 - 836\,\kappa + 103\,\kappa^2 - 26\,\kappa^3
              \right) \,c( \,u)
           \right) \,{u}^3 - \nonumber \\
& \left. \left. {\sqrt{8\,\kappa -6}}\,
         \left( 1519 - 596\, \kappa + 93\, \kappa^2 +
           \left( -977 + 432\,\kappa + 65\,\kappa^2 \right) \,
            c( \,u) \right) \,{u}^4
         + \right. \right. \nonumber \\
&  \left. \left.  \left( 2265 - 2612 \kappa + 213 \kappa^2 + 34
     \kappa^3 -  \left( 385 - 600 \kappa + 141 \kappa^2 + 14 \kappa^3
              \right) \,c( \,u)
           \right) {u}^5 + \right. \right. \nonumber \\
&   \left. \left.    {\sqrt{8\,\kappa -6}}\,
         \left( -41 + 84\,\kappa - 43\,\kappa^2 +
           \left( 109 + 48\,\kappa - 69\,\kappa^2 \right) \,
            c( \,u) \right) \,{u}^6
         + \right. \right.  \nonumber \\
&  \left. \left.       \left( 267 - 60\,\kappa - 305\,\kappa^2 +
             54\,\kappa^3 -
           2\,\left( 485 - 872\,\kappa + 361\,\kappa^2 -
              18\,\kappa^3 \right) \,
            c( \,u) \right) \,{u}^7
         + \right. \right. \nonumber \\
&  \left. \left.     4\,{\sqrt{8\,\kappa -6}}\,
         \left( 99 - 112\,\kappa + 21\,\kappa^2 \right) \,{u}^8 \right) \,{u'(r)}^2\right)
         \left(2\, \,r^2\,\kappa\,
     {c( \,u)}^3\, {d}( \,u)\,
     \left( 8\,\left( 6 - \kappa + \right. \right. \right. \nonumber \\
&  \left. \left. \left.   2\,\kappa^2 +
          \left( -6 - \kappa + 2\,\kappa^2 \right) \,
           c( \,u) \right)  +
       8\, \,{\sqrt{8\,\kappa -6}}\,
        \left( 2 + \kappa + \left( \kappa - 2 \right) \,
           c( \,u) \right) \,u + \right. \right. \nonumber \\
& \left. \left.  4\,\left( -4 + 5\,\kappa +
          2\,\left( \kappa - 2 \right) \,\kappa\,
           c( \,u) \right) \,{u}^2
        + 2\,{\sqrt{8\,\kappa -6}}\,
        \left( 3\,\kappa + 5\,\left( \kappa - 2 \right) \,
           c( \,u) \right) \,{u}^3
        + \nonumber \right. \right. \\
&  \left. \left.    \left( -49 + 57\,\kappa - 6\,\kappa^2 \right)
\,{u}^4
       \right)\right)^{-1} = 0
\end{align}
\begin{align}
r \frac{d^2}{dr^2} \left( r \phi_1 \right) - \left( \kappa \,
{\left(\left( 3\,c(u)- 1 \right) -  {\sqrt{ 8\,\kappa - 6}}\,u
\right) }^2 \phi_1\right) d(u)^{-1}  = 0
 \end{align}
where the functions $c(u)$ and $d(u)$ are defined in equations
\eqref{def-c} and \eqref{def-d} respectively.



\end{document}